# A Lightweight and Attack Resistant Authenticated Routing Protocol for Mobile Adhoc Networks

M.Rajesh Babu[1] and S.Selvan[2]

[1]Department of Computer Science and Engineering,
PSG College of Technology, Coimbatore, TamilNadu, India
mrb@cse.psgtech.ac.in

[2]Principal, Francis Xavier Engineering College,
Tirunelveli, TamilNadu, India
drselvan@ieee.org

## ABSTRACT

In mobile ad hoc networks, by attacking the corresponding routing protocol, an attacker can easily disturb the operations of the network. For ad hoc networks, till now many secured routing protocols have been proposed which contains some disadvantages. Therefore security in ad hoc networks is a controversial area till now. In this paper, we proposed a Lightweight and Attack Resistant Authenticated Routing Protocol (LARARP) for mobile ad hoc networks. For the route discovery attacks in MANET routing protocols, our protocol gives an effective security. It supports the node to drop the invalid packets earlier by detecting the malicious nodes quickly by verifying the digital signatures of all the intermediate nodes. It punishes the misbehaving nodes by decrementing a credit counter and rewards the well behaving nodes by incrementing the credit counter. Thus it prevents uncompromised nodes from attacking the routes with malicious or compromised nodes. It is also used to prevent the denial-of-service (DoS) attacks. The efficiency and effectiveness of LARARP are verified through the detailed simulation studies.

## 1. INTRODUCTION

A mobile ad-hoc network (MANET) is a multi-hop wireless network, a temporary and without infrastructure in which the nodes can move randomly. These MANETS are able to extend their wireless transmission range of each node by multi-hop packet forwarding. So these MANETS are suited for the situations in which pre deployed infrastructure support is not available. An ad hoc network doesn't have any fixed infrastructure like base stations or mobile switching centers. Mobile nodes which are within the radio range to each other can communicate directly through wireless links, whereas the nodes which are far away depend on other nodes to communicate messages as routers. In an ad hoc network the node mobility causes frequent changes of the network topology. Mobile ad hoc networks have their applications in both military and civilian circumstances due to their self-organizing and self-configuring potentials.

The routing aspects of MANETs are discussed earlier, while the research activities about security in MANETs are in their beginning stage. Apart from the regular network problems MANETs creates new security problems. Ad hoc networks use all the available nodes for routing and forwarding to increase the throughput of the total network. Therefore when more

 

nodes participate in packet routing, it increases the total bandwidth and decreases the possible routing paths and also the possibility of network partition.

When a node is overloaded, selfish, malicious or broken it may misbehave by not approving to forward packets. An overloaded node does not have the CPU cycles, buffer space or available network bandwidth to forward packets. A selfish node expects other nodes to forward packets because it is not willing and does not have direct interest to spend its battery life, CPU cycles or available network bandwidth to forward packets. A malicious node introduces a denial of service attack by dropping packets. A broken node prevents it from forwarding packets by having a software fault.

The mobile ad hoc network needs more security mechanisms than in fixed networks. Through the compromised nodes attackers can interrupt into the network. When the nodes join or leave the network, and roam in the network often, then the network topology becomes highly dynamic. Mobile users request security services when they move from one place to another due to its dynamic nature. To achieve protection and high network performance, a powerful security solution is needed, otherwise

- The security system may misbehave when an attacker infiltrates the security system.

- The network performance may be degraded by the misbehavior of the nodes.

- The malicious nodes can work as routers and stops the network from delivering the packets correctly when they are not protected properly.

For example, the malicious node can declare faulty routing updates which are spread through all over the network or it may drop the packets which are passing across them. A very important security issue in ad hoc network is to protect the operations of their network layer from malicious attacks. In a distributed computer systems, there are lot of familiar attacks which include

- Denial of Service: Due to overload or malfunction the network service is not available

- Information theft:  Through an illegal instance, information is read.

- Intrusion: Through an unauthorized person, access is achieved to some restricted service.

- Tampering: Through an unauthorized person, the modification is achieved.

In this paper, we propose to develop a Lightweight and Attack Resistant Authenticated Routing Protocol (LARARP) for mobile ad hoc networks, which will mitigate the routing misbehavior of nodes in mobile ad hoc networks.

The protocol involves:

- The destination node authenticates the source node

- The destination node authenticates every intermediate node listed in the packet header

- The source and destination node confirm the correctness of node's sequence in the node list

- Well behaving nodes are rewarded by more credits and misbehaving nodes are punished by reducing the credits

## 2. RELATED WORK

Farooq Anjum et al. [1] proposed an initial approach to detect the intrusions in ad hoc networks. In this work, the network activities are observed by the signature based IDS and compares them with the known attacks. The new unknown threats cannot be detected is the limitation of this approach.

Anand Patwardhan et al. [2] proposed a secure routing protocol based on AODV over IPv6. In additional to this, by the routing protocol independent Intrusion Detection and Response system for ad hoc networks it is reinforced.

Yih-Chun Hu et al [3] have designed and estimated a secure ad hoc network protocol called Secure Efficient Ad hoc Distance vector routing protocol (SEAD). It is based on the design of the Destination-sequence Distance-vector routing protocol (DSDV). They have used the efficient one way hash functions without using an asymmetric cryptographic operation in the protocols in order to support the node usage with limited CPU processing capability and also to protect against the Denial of Service (DoS) attacks where the attacker attempts to cause other nodes to use more network bandwidth or processing time. SEAD are tested in various ranges of situations and performed well. For the multiple uncoordinated attackers creating incorrect routing state in other nodes, and also in the active attackers or compromised nodes in the network, SEAD is robust.

Tarag Fahad et al. [4] have proposed a new mechanism, Packet Conservation Monitoring Algorithm (PCMA) by focusing on the detection phase. This is used to detect the selfish nodes in MANETs. It does not show other threats because these protocols address the issue of packet forwarding attacks.

Yih Chun Hu et al. [5] have presented the attacks against routing in ad hoc networks and estimated the performance of a secure on demand ad hoc network routing protocol called Ariadne. The corruption of the uncompromised routes which consists of the uncompromised nodes by the attackers or compromised nodes was prevented by Ariadne. It is also used to prevent the Denial of Service attacks. Moreover, Ariadne is effective and uses only the highly efficient symmetric cryptographic primitives.

Panagiotis Papadimitratos and Zygmunt J. Haas [6] have discussed about the Secure Routing Protocol (SRP) counters. Its malicious behavior brings the discovery of the topological information. An irregularly misbehaving attacker is first conforming to the route discovery to make itself as a part of route, and then corrupt the in-transit data. With the help of the Secure Message Transmission Protocol (SMT), protection of data transmission has been given. This provides a flexible, end to end secure data forwarding scheme which naturally complements SRP. The processing overhead due to the cryptographic operations remains low. This allows the protocol to remain competitive to reactive protocols which do not incorporate the security features.

Yanchao Zhang et al [7] have proposed a credit based Secure Incentive Protocol (SIP) for simulating the cooperation in packet forwarding for MANETs without infrastructure. It does not show other threats because these protocols address the issue of packet forwarding attacks.

Liu et al. [8] have proposed a 2ACK scheme which works as an add-on technique for routing methods for detecting the routing misbehavior and to reduce the adverse effect. The unnecessary overhead is resulted by sending the acknowledgement packets even if there is no misbehavior.
Li Zhao et al. [9] have proposed a new method MARS and its enhancement E-MARS for detecting the misbehavior and to reduce the adverse effect in adhoc networks, they. In this method, the information packets are prevented to reach the destination by a route failure or link failure. In additional to this, the destination may not be able to detect the misbehavior when a selfish node does not forward the information packet or modifies the contents of the information packets.

Patwardhan et al. [10] have presented their method in securing a MANET with the help of the threshold-based intrusion detection system and a secure routing protocol. A proof of concept execution has been presented for their IDS organized on handheld devices and also in a MANET testbed which is connected by a secured version of AODV over IPv6-SecAODV.The attacks on the data traffics are detected by the IDS. SecAODV is used for the security features of non-repudiation and authentication without depending upon the Certificate Authority (CA) or Key Distribution Center (KDC) availability. The design and implementation details of their system, the practical considerations involved and the working of these mechanisms which are used to detect and prevent the malicious attacks are presented by them.

Huaizhi Li and Mukesh Singhal [11] have presented an on-demand secure routing protocol for ad hoc networks which is based on a distributed authentication mechanism. In order to establish a trust relationship between the network entities, this protocol has made use of the recommendation and trust evaluation and also it uses the feedback to adjust it. Without the third party support it discovers the multiple routes between two nodes.

Sergio Marti, et al., has explained two techniques which improves the throughput in an adhoc network. In order to reduce the effects of routing misbehavior in adhoc networks such as watchdog and path rater, they have examined two possible extensions to DSR. The misbehaving nodes are identified by the watchdog and the path rater assists the routing protocol to avoid the misbehaving nodes [12].

Katrin Hoeper and Guang Gong have introduced two full functional identity-based authentication and key exchange methods for mobile adhoc networks. For IBC (Identity-based cryptographic) schemes they have presented the first key revocation and key renewing algorithms. For designing MANET-IDAKE schemes they have used some features of IBC schemes like pre shared secret keys from the pairings and efficient key management which meets the special constraints and the requirements of MANETs [13].

Gergely Acs et, al. [14] have proposed a mathematical framework in which security is defined exactly and it has proved that the routing protocols for mobile ad hoc networks to be secure exactly. For the on demand source routing protocols they have modified their framework but the general principles are applicable to the other protocols. Their approach is used based on the simulation paradigm which is widely used to examine the key establishment protocols. But still it is not applied in the situations like ad hoc routing. They proposed an on demand source routing protocol called as endairA. They have also proved that their framework is secured and established it's utility.

Syed Rehan Afzal et al. [15] have determined the security problems and attacks in existing routing protocols. In order to remove the problems which are mentioned in the existing protocols, they have presented the design and analysis of a secure on demand routing protocol called RSRP. In addition, RSRP is different than Ariadne and uses an effective broadcast authentication mechanism without clock synchronization and helps the instant authentication.

# 3. SYSTEM DESIGN AND ALGORITHM OVERFLOW

In this paper, we propose a Lightweight and Attack Resistant Authenticated Routing protocol (LARARP). An effective security is given by our protocol for the route discovery attacks in the MANET routing protocols. It helps the nodes to drop the invalid packets earlier by detecting the malicious node quickly.

In our proposed protocol, before transmitting the data to the destination the sender generates a temporary key pair. A secret key list SS and public key PS are present in the key pair. In order to generate the secret list SS0, SS1, it uses the concept of one-way hash function. The public key PS can be created by hashing every element of the SSi. The public key PS is sent to the respective destinations by the sender after this key generation. Then the verification information is created by the source using the SS list. The verification information and broadcasted in the route request packet. Each intermediate node will check the verification of the source by receiving the route request packet, using its PS. The packet will be discarded if the verification fails else it will be forwarded.

The destination will check the validity of the verification of the source when the route request reaches the destination. It discards the packet if it fails.

A MAC based authentication code is used for the reliability of route request packet. If there is any modification in the content of the route request packet including verification information by the intermediate node, then it is detected by the destination node by checking the MAC code using its public key. It discards the packet if it gets modified. The route reply packet is transmitted by the destination node. The same steps of the route request packet are repeated for the reply packet.

In addition to this, our proposed protocol presents a credit based incentive scheme to punish and reward the misbehaving and well behaving nodes respectively. It is a natural idea to encourage cooperation among selfish nodes by rewarding them with some credits. Only when a node has enough credits in hand, it could transmit its own packets.

We have taken the AODV routing protocol as the base and modified according to the proposed protocol.

## 3.1. General Attacks on Ad Hoc Network Routing Protocols

Ad hoc network routing protocol attack usually lies in any of these two categories: routing disruption attacks and resource consumption attacks. The attacker makes effort to cause reasonable data packets to be routed in dysfunctional ways in the routing disruption attack. The attacker inserts packets into the network in order to consume precious network resources bandwidth, or node resources such as memory (storage) or computation power in the resource consumption attack. Both attacks are the example for a Denial-of-service (DoS) attack by viewing through an application layer.

If the attacker sends fake routing packets to create routing loop which causes the packets to cross the nodes in a cycle without reaching their target, consuming energy and available bandwidth is the best example of a routing disruption attack.

Similarly the attacker can also create a routing black hole; the packets are dropped in this by sending fake routing packets. The attacker can also route all packets for some destination to itself and then it can discard them or the attacker can also make of all the nodes in an area of the network to point "into" that area even if the destination is outside the area.

An attacker may create a gray hole as a special case of black hole in which it drops only some selected packets. For example, forwarding the routing packets but not the data packets. An attacker may also take effort to cause a node to use diversion (suboptimal routes) or by inserting the fake routing packets to makes effort to divide the network in order to prevent one set of nodes from reaching the other.

If the attacker makes effort to create a route to look longer by itself through adding virtual nodes to the route, then this attack is called as gratuitous detour where the shorter route does not exist.

The rushing attack is a malicious attack which is targeted against the on-demand protocol which uses duplicate suppression at each node. Throughout the network the attacker distributes ROUTE REQUESTS quickly by holding back any reasonable ROUTE REQUESTS, while the nodes drop them due to duplicate suppression.

If an attacker injects extra data packets to the networks which use bandwidth resources when forwarded particularly to over charge or routing loops, is the best example of a resource consumption attack.

Likewise the attacker can also inject extra control packets to the network which uses still more bandwidth or computational resources in which the other nodes process and forward such packets. An Active-VC attacker can try to take out more resources from the nodes on both sides the vertex cut by using either of these attacks. For example, the nodes wastes the energy by forwarding packets to the vertex cut which is to be dropped is achieved  through forwarding only the routing packets but not the data packets.

### 3.2. Incentive Scheme

An additional data structure called Neighbor's Trust Counter Table (NTT) is maintained by each network node. Let {CC1, CC2,…} be the initial trust counters of the nodes {N1, N2,…} along the route from a source S to the destination D.

Since the node does not have any information about the reliability of its neighbors in the beginning, nodes can neither be fully trusted nor be fully distrusted. When a source S wants to send a packet to the destination D, it sends route request (RREQ) packets.

Each node keeps track of the number of packets it has forwarded through a route using a credit counter (CC).   Each time, when node Nk receives a packet from a node Ni, then Nk increases the credit counter of node Ni as

$$CCNi = CCNi + 1, i =1, 2……$$ (1)

Then the NTT of node Nk is modified with the values of CCNi. Similarly each node determines its NTT and finally the packets reach the destination D. When the subsequent RREQ message reaches the destination, it checks the credit values of the intermediate nodes, before verifying their digital signature. The nodes are considered as well behaving nodes if the credit values are equal or greater than a credit threshold Ct. On the other hand, the nodes are considered as misbehaving nodes if the credit values are less than Ct. The verifications of the digital signature are made only to the misbehaving nodes instead of verifying all the nodes, which reduces the control overhead. Also nodes with credit counter values less than Ct are prohibited from further transmissions.

The source will be considered as a route breakage or failure, if it does not receives the RREP packet for a time period of t seconds. Then the route discovery process is started by the source again.

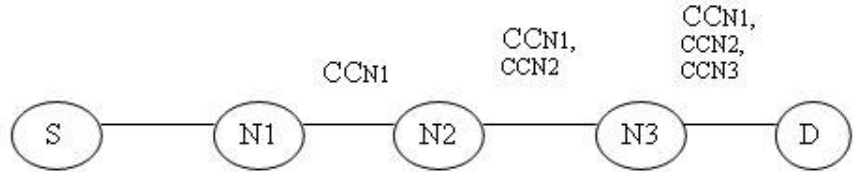

Figure 1:  Incentive Scheme

### 3.3. Route Discovery Process

In the proposed protocol, once a node S want to send a packet to a destination node D, it initiates the route discovery process by constructing a route request RREQ packet. It contains the source and destination ids and a request id, which is generated randomly and a MAC computed over the request id with a key shared by the sender and the destination.

When an intermediate node receives the RREQ packet for the first time, it appends its id to the list of node ids and signs it with a key which is shared with the destination. It then forwards the RREQ to its neighbors.

Let N1, N2….Nm-1 nodes are there, between the source S and the destination D.

The route request process is illustrated as below:

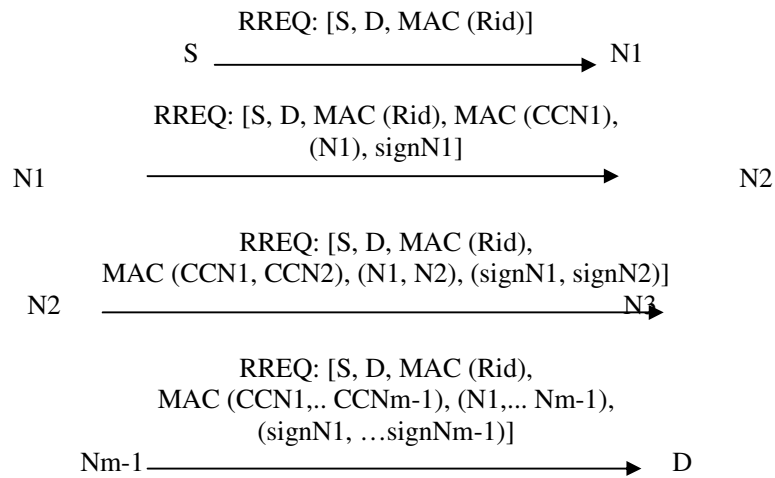

Figure 2:  Route Request Process

When the destination receives the accumulated RREQ message, it first verifies the credit values of the intermediate nodes and then the sender's request id by recomputing the sender's MAC value, with its shared key. It then verifies the digital signature of each intermediate node. With the credit values of the intermediate nodes, the threshold Ct is compared to detect the misbehaving node. If all these verifications are successful, then the destination generates a route reply message RREP. If the verifications fail, then the RREQ is discarded by the destination. It again constructs a MAC on the request id with the key shared by the sender and the destination. The RREP contains the source and destination ids, The MAC of the request id, the accumulated route from the RREQ, which are digitally signed by the destination. The RREP is sent towards the source on the reverse route.

When the intermediate node receives the RREP packet, it checks whether its id is in the list of ids stored by the RREP. It also checks for the ids of its neighbors in the list.

The intermediate node then verifies the digital signature of the destination node stored in the RREP packet, is valid. If the verification fails, then the RREP packet is dropped. Otherwise, it is signed by the intermediate node and forwarded to the next node in the reverse route.

When the source receives the RREP packet, if first verifies that the first id of the route stored by the RREP is its neighbor. If it is true, then it verifies all the digital signatures of the intermediate nodes, in the RREP packet. If all these verifications are successful, then the source accepts the route. The source also verifies the request id that it sent along with RREQ packet. If it received back the same request id from the destination, it means that there is no replay attack.

The route reply process is illustrated as below:

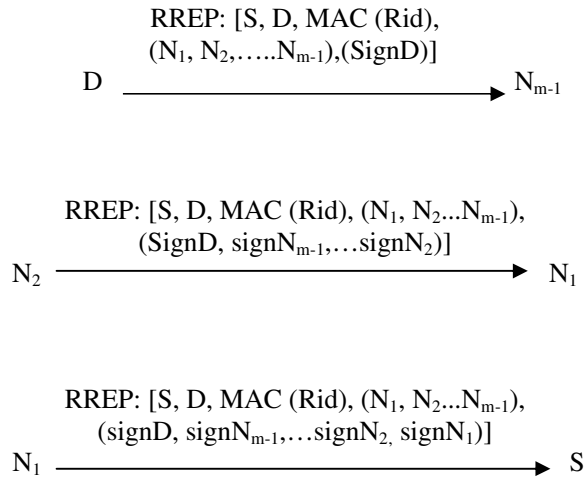

Figure 3: Route Reply Process

In this protocol, authentication is performed for both route request and route reply operations. Also, only nodes which are stored in the current route need to perform these cryptographic computations. So the proposed protocol is efficient and more secure.

## 4. PERFORMANCE EVALUATION

### 4.1 Simulation Model and Parameters

We use NS2 to simulate our proposed algorithm. In our simulation, the channel capacity of mobile hosts is set to the same value: 2 Mbps. We use the distributed coordination function (DCF) of IEEE 802.11 for wireless LANs as the MAC layer protocol. It has the functionality to notify the network layer about link breakage.

In our simulation, 100 mobile nodes move in a 1000 meter x 1000 meter square region for 50 seconds simulation time. We assume each node moves independently with the same average speed. All nodes have the same transmission range of 250 meters. In our simulation, the minimal speed is 5 m/s and maximal speed is 10 m/s. The simulated traffic is Constant Bit Rate (CBR).

Our simulation settings and parameters are summarized in the following table:

Table 1: Default Simulation Parameters

| Parameters | Assumptions |
|---|---|
| No. of Nodes | 100 |
| Area Size | 1000 X 1000 |
| Mac | 802.11 |
| Radio Range | 250m |
| Simulation Time | 50 sec |
| Traffic Source | CBR |
| Packet Size | 512 |
| Speed | 5m/s t 10m/s |
| Misbehaving Nodes | 5,10,15,20, 25 |
| Pause time | 10,20,30,40, 50 |

### 4.2 Performance Metrics

We evaluate mainly the performance according to the following metrics.

Control overhead: The control overhead is defined as the total number of routing control packets normalized by the total number of received data packets.

Average end-to-end delay: The end-to-end-delay is averaged over all surviving data packets from the sources to the destinations.

Average Packet Delivery Ratio: It is the ratio of the number of packets received successfully and the total number of packets transmitted.

The simulation results are presented in the next section. We compare our LARARP with the SAODV [14] protocol in presence of malicious node environment.

## 5. RESULTS

### 5.1 Based on Malicious nodes

In our First experiment, we vary the no. of misbehaving nodes as 5,10,15,20 and 25.

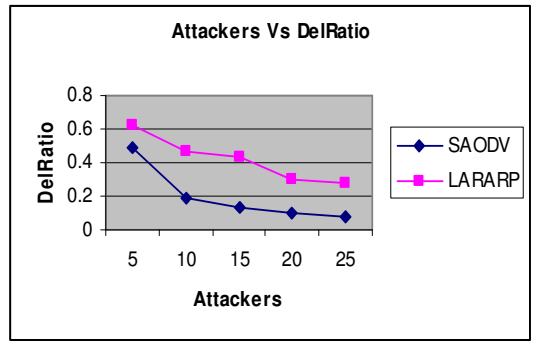

Figure 4 Attackers Vs Delivery Ratio

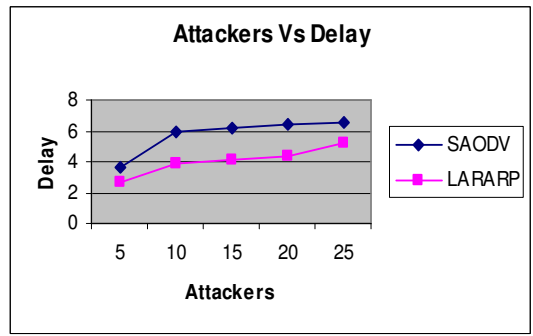

Figure 5 Attackers Vs Delay

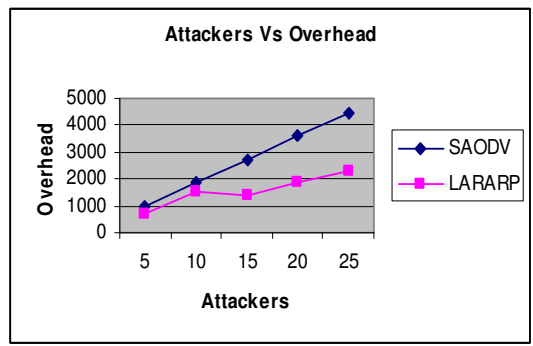

Figure 6 Attackers Vs Overhead

Figure 4 shows the results of average packet delivery ratio for the misbehaving nodes 5, 10….25 for the 100 nodes scenario. Clearly our LARARP scheme achieves more delivery ratio than the SAODV scheme since it has both reliability and security features.

Figure 5 shows the results of average end-to-end delay for the misbehaving nodes 5, 10… 25. From the results, we can see that LARARP scheme has slightly lower delay than the SAODV scheme because of authentication routines.

Figure 6 shows the results of routing overhead for the misbehaving nodes 5, 10….25. From the results, we can see that LARARP scheme has less routing overhead than the SAODV scheme since involves route re-discovery routines.

## 5.2 Based on Pause Time

In our Second experiment, we vary the pause time as 10,20,30,40 and 50, with 5 attackers.

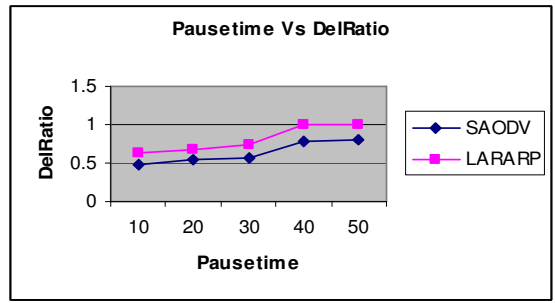

Figure 7 Pause Time Vs Delivery Ratio

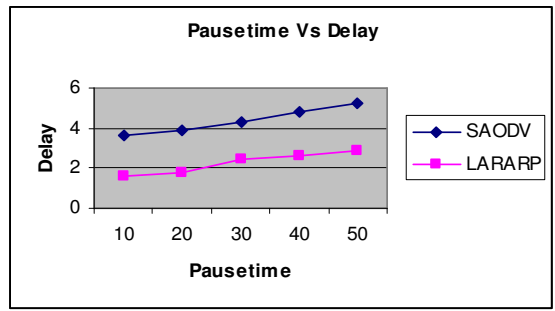

Figure 8   Pause Time Vs Delay

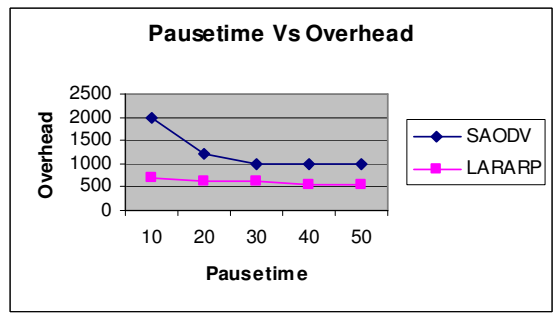

Figure 9   Pause Time Vs Overhead

Figure 7 shows the results of average packet delivery ratio for the pausetimes 10, 20…50 for the 100 nodes scenario. Clearly our LARARP scheme achieves more delivery ratio than the SAODV scheme since it has both reliability and security features.

Figure 8 shows the results of average end-to-end delay for the pausetimes 10, 20….50. From the results, we can see that LARARP scheme has slightly lower delay than the SAODV scheme because of authentication routines.

Figure 9 shows the results of routing overhead for the pausetimes 10, 20….50. From the results, we can see that LARARP scheme has less routing overhead than the SAODV scheme since involves route re-discovery routines.

## 6. CONCLUSIONS

This paper has presented the design and evaluation of a (LARARP), a new ad hoc network routing protocol that provides security against routing misbehaviors and attacks. It relies on cryptographic techniques for authentication. The design is based on the AODV routing protocol. Our protocol supports the nodes to drop the invalid packets earlier by detecting the malicious nodes quickly by verifying the digital signatures of all the intermediate nodes. It punishes the misbehaving nodes by decrementing a credit counter and rewards the well behaving nodes by incrementing the credit counter. Thus it prevents uncompromised nodes from attacking the routes with malicious or compromised nodes It also prevents a variety of denial-of-service (DoS) attacks. Detailed simulation studies have confirmed the efficiency and effectiveness of LARARP. By comparing with the existing scheme, we have shown that the proposed protocol attains high packet delivery ratio with reduce delay and overhead.

## ACKNOWLEDGEMENTS

We would like to thank the anonymous reviewers, whose comments and suggestions helped to improve the presentation of this paper.

## REFERENCES

[1]    Farooq Anjum and Dhanant Subhadrabandhu and Saswati Sarkar  "Signature based  Intrusion Detection for Wireless Ad-Hoc Networks:  A Comparative study of various routing protocols" Vehicular Technology Conference, 2003. VTC 2003-Fall. 2003 IEEE 58th, Oct. 2003.

[2]    Anand Patwardhan, Jim Parker, Anupam Joshi, Michaela Iorga and Tom Karygiannis "Secure Routing and Intrusion Detection in Ad Hoc Networks" Pervasive Computing and Communications, 2005. PerCom 2005. Third IEEE International Conference on March 2005.

[3]    Yih-Chun Hu, David B. Johnson and Adrian Perrig, "SEAD: Secure    Efficient Distance Vector Routing for Mobile Wireless Ad Hoc  Networks", in  proceedings of IEEE Workshop on Mobile Computing  Systems and  Applications, pp.3-13, 2002.

[4]    Tarag Fahad & Robert Askwith "A Node Misbehaviour Detection Mechanism for Mobile Ad-hoc Networks" The 7th Annual PostGraduate Symposium on The Convergence of Telecommunications, Networking and Broadcasting, 26-27 June 2006.

[5]    YihChun Hu, Adrian Perrig and David B. Johnson," Ariadne: A Secure on  Demand Routing Protocol for Ad Hoc Networks", Technical Report, Rice university 2001.

[6]      Panagiotis Papadimitratos and Zygmunt Haas, "Secure Routing for     Mobile Ad Hoc Networks", in proceedings of conference on SCS Communication Networks and Distributed Systems Modeling and Simulation, pp.27-31,  2002.

[7]      Yanchao Zhang, Wenjing Lou, Wei Liu, and Yuguang Fang, "A secure  incentive protocol for mobile ad hoc networks", Wireless Networks (WINET), vol 13, issue 5, October 2007.

[8]      Liu, Kejun Deng, Jing Varshney, Pramod K. Balakrishnan, Kashyap "An   Acknowledgment-based Approach for the Detection of Routing Misbehavior in MANETs" Mobile Computing, IEEE Transactions on  May 2007.

[9]      Li Zhao and José G. Delgado-Frias "MARS: Misbehavior Detection in  AdHoc Networks" Global Telecommunications Conference, 2007.  GLOBECOM '07. IEEE Publication Date: 26-30 Nov. 2007.

[10]      A. Patwardhan,J. Parker, M. Iorga, A. Joshi,T. Karygiannis and Y. Yesha, "Threshold-based intrusion detection in ad hoc networks and secure  AODV", Vol.6, No.4, pp.578-599, 2008.

[11]      Huaizhi Li and Mukesh Singhal, "A Secure Routing Protocol for    Wireless Ad Hoc Networks", in proceedings of 39th Annual Hawaii International  Conference on System Sciences, Vol.9, 2006.

[12]      Sergio Marti, T.J. Giuli, Kevin Lai, and Mary Baker," Mitigating Routing  Misbehavior in Mobile Ad Hoc Networks", in proc. of 6th International Conference on Mobile computing and networking,pp:255- 265, May  2000.

[13]      Katrin Hoeper and Guang Gong, "Bootstrapping Security in Mobile Ad Hoc Networks Using Identity-Based Schemes with Key Revocation",  Technical Report CACR 2006-04, Centre for Applied Cryptographic Research, January 2006.

[14]      Gergely Acs, Levente Buttya, and Istvan Vajda, "Provably Secure On-Demand Source Routing in Mobile Ad Hoc Networks", IEEE Transactions On Mobile Computing, Vol. 5, No. 11, pp.1533-1546, November 2006.

[15]      Syed Rehan Afzal, Subir Biswas, Jong-bin Koh, Taqi Raza, Gunhee Lee, and Dong-kyoo Kim, "RSRP: A Robust Secure Routing Protocol for  Mobile Ad hoc Networks", IEEE Conference on Wireless Communications and Networking, pp.2313-2318, April 2008.

**Authors**

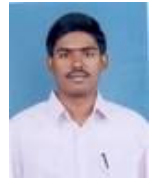

M.Rajesh Babu received the B.E. degree in computer science and engineering from the Bangalore University, Bangalore, India in 2000 and the M.E. degree in computer science and engineering from the Anna University, Chennai, India in 2004 and currently pursuing the Ph.D. degree in information and communication engineering from Anna University, Chennai, India. He has 8 years of teaching and industry experience. He is currently working as a Senior Lecturer of computer science and engineering at PSG college of Technology, Coimbatore, India. He is a member of SSI. He has published 5 papers in international journals 5 in national conference proceedings. His areas of interest include computer networks, object oriented analysis and design, network security and software engineering.

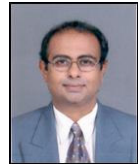

Dr. S.Selvan received the B.E. degree in electronics and communication engineering and the M.E. degree in communication systems from the University of Madras, Chennai, India, in 1977 and 1979, respectively, and the Ph.D. degree in computer science and engineering from the Madurai Kamaraj University, Madurai, India, in 2001. He has 30 years of teaching and research experience. He is currently working as Principal at Francis Xavier engineering college, tirunelveli, India. He is a senior member of IEEE and fellow of IE(I) and IETE. He has published more than 120 papers in international and national journals and conference proceedings. His areas of research include data mining, soft computing, computer networks, signal processing, image processing and network security.